\documentclass[10pt,a4paper]{article}

\usepackage[latin1]{inputenc}
\usepackage[francais,english]{babel}
\usepackage{graphicx}
\begin{document}
\twocolumn[

\title{ 
\begin{minipage}[t]{170mm}
\begin{center} \bf
Precision measurement of the half-life and the decay branches of $^{62}$Ga
\end{center}
\end{minipage}
}

\author{
\begin{minipage}[t]{170mm}
\begin{center}
G. Canchel$^1$, B. Blank$^1$, M. Chartier$^2$, 
F. Delalee$^1$, P. Dendooven$^3$, C. Dossat$^1$, J. Giovinazzo$^1$, 
J. Huikari$^4$, A. S. Lalleman$^1${\footnotemark[2]},
M. J. Lopez Jiménez$^1$, V. Madec$^1$, J. L. Pedroza$^1$, H. Penttil\"a$^4$,
J. C. Thomas$^1${\footnotemark[3]}  
\end{center}
\end{minipage}
}
\maketitle

$^1$ Centre d'\'Etudes Nucl\'eaires de Bordeaux-Gradignan, Le Haut-Vigneau, 
F-33175 Gradignan Cedex, France\\
$^2$ Oliver Lodge Laboratory, University of Liverpool, Liverpool L69 7ZE, UK\\
$^3$ Kernfysisch Versneller Instituut, Zernikelaan 25, NL 9747 AA Groningen, 
Netherlands\\
$^4$ Department of Physics, University of Jyv\"askyl\"a,
P.O. Box 35, FIN-40351 Jyv\"askyl\"a, Finland

\begin{abstract}
\noindent 
In an experiment performed at the Accelerator Laboratory of the University of 
Jyv\"askyl\"a, the $\beta$-decay half-life of $^{62}$Ga has been studied with
high precision using the IGISOL technique. A half-life of T$_{1/2}$ = 116.09(17)~ms 
was measured. Using $\beta$-$\gamma$ coincidences, the $\gamma$ intensity of the 954~keV 
transition and an upper limit of the $\beta$-decay feeding of the 0$^+_2$ state have 
been extracted. The present experimental results are compared to previous measurements
and their impact on our understanding of the weak interaction is discussed.
\end{abstract}
\vspace*{5mm}
]

\renewcommand{\footnotesep}{0cm}
\renewcommand{\footnoterule}{\rule{7.5cm}{0.01cm}\vspace{0.2cm}}
\renewcommand{\thefootnote}{\fnsymbol{footnote}}
\footnotetext[2]{present address: CEA, 91191 Gif-sur-Yvette, France}
\footnotetext[3]{present address: Grand acc\'el\'erateur national d'ions lourds,
                 BP~55027, F-14076 Caen Cedex 5, France}

\section{Introduction}
  
Studies  of superallowed 0$^+$ $ \rightarrow$ 0$^+$ pure Fermi transitions and the precise 
determination of their {\it ft} value are important as tests for the weak interaction 
theory. The corrected {\it Ft}-value of these transitions serves as a stringent 
test for the Conserved Vector Current (CVC) hypothesis, for the determination of the 
vector coupling constant G$_{\nu}$ of nuclear $\beta$-decay and for the validity of 
the three-generation Standard Model through the determination of the {\it V$_{ud}$} 
element of the Cabbibo-Kobayashi-Maskawa (CKM) quark mixing matrix. The following relation
links the coupling constant G$_{\nu}$ to the corrected {\it Ft}-value~\cite{Tow98}:

\begin{equation}
    Ft=ft(1+\delta_R')(1+\delta_{NS}-\delta_C)=\frac{k}{2G_v^2(1+\Delta_R^v)}
\end{equation}

Here, {\it f} is the $\beta$-decay phase space factor, {\it t} is the partial half-life, 
{\it k} is a constant ( {\it k/($\hbar$c)$^6$} = (8120.271 $\pm$ 0.012)$ \times 
$10$^{-10}$ GeV$^{-4}$s )~\cite{Tow98}, $\delta_R$, $\delta_{NS}$ and $\Delta_R^v$ are radiative 
corrections and $\delta_C$ accounts for the deviation of the parent and daughter states 
from perfect isospin symmetry. Presently, nine transitions have been measured to a 
precision equivalent or better than 10$^{-3}$ : $^{10}$C, $^{14}$O, $^{26}$Al$^m$, 
$^{34}$Cl, $^{38}$K$^m$, $^{42}$Sc, $^{46}$V, $^{50}$Mn and $^{54}$Co~\cite{Tow98,Har90}. 
The data concerning these transitions confirm the CVC hypothesis at the 3$\times$10$^{-4}$ 
precision level but until very recently the unitarity of the first row of the 
CKM matrix was not confirmed at more than 2.2 standard deviations~\cite{Har90,Tow02}. 
New measurements seem to indicate that the value of the up and strange mixing matrix 
element (V$_{us}$) was too low~\cite{She03}. The new value of V$_{us}$ could 
indicate that unitarity may be completely restored ($\Sigma$V$_{ij}^2$=0.9999(16) 
instead of 0.9968(14) before). However, more measurements are certainly needed to
confirm this finding.

Deviation from unitarity would have important consequences for the description 
of the weak interaction through the Standard Model. However, before claiming new physics 
beyond the standard model, a very careful determination of the corrected {\it Ft}-value is needed, 
where especially the isospin symmetry breaking factor $\delta_C$ is an important input. 
In fact, different model calculations predict that this correction could strongly 
increase for heavier N=Z, odd-odd nuclei. Therefore measurements for these 
heavier nuclei are needed to test the evolution of the correction and thus to check
nuclear models used for the calculations.

The aim of this work is a precise determination of the half-life of $^{62}$Ga
as well as of its decay branches which are part  
of the inputs necessary to calculate the corrected {\it Ft} value for $^{62}$Ga. 
This experiment constitutes part of the efforts~\cite{Bla02} at IGISOL in Jyv\"askyl\"a 
to measure the half-life and the $\beta$-decay branching ratios as well as, 
using the new IGISOL ion trap, the masses of $^{62}$Zn and $^{62}$Ga 
in order to extract the Q value of the reaction.

The half-life has been measured several times~\cite{Alb78,Chi78,Dav79,Hym03,Bla04} in the past, 
however, except for the very recent experiment performed at GSI~\cite{Bla04}, the 
precision on the result was too low to be included in the CVC test and the determination 
of the CKM matrix unitarity. The results presented in this paper achieve the necessary 
precision concerning the half-life.

\section{Experimental setup}

The $^{62}$Ga activity was produced via the $^{64}$Zn(p,3n) fusion-evaporation reaction. 
The $\beta$ decay of $^{62}$Ga was studied after mass separation using the IGISOL 
technique~\cite{Den97,Ays01} which is independent of the chemical and physical 
properties of the elements and well suited for short-lived activities.
The 3~mg/cm$^2$ $^{64}$Zn target used as the entrance window of the 
light-ion ion guide was bombarded with a 48 MeV proton beam (20~$\mu$A) from the k = 130 
Jyväskylä cyclotron.

The 40 keV mass-separated A=62 beam was impinging on a 1/2 inch wide movable tape at 
the end of the central IGISOL beam line. The tape was located at the center of a 
4$\pi$ cylindrical plastic scintillator (2 mm thick, 12~mm diameter entrance hole) 
which was used for $\beta$ 
detection. The scintillation light was collected by two 2-inch photomultiplier 
tubes in coincidence through a special light guide. The coincidence requirement 
allowed us to remove most of the photomultiplier noise. The $\beta$-particle 
detection efficiency was 90 \%. Three co-axial germanium detectors (40\% 
relative efficiency) in close geometry were used 
in order to provide $\beta$-$\gamma$ and $\beta$-$\gamma$-$\gamma$ coincidence data.
They were mounted at a typical distance of 40 mm from the source point and had 
a total photopeak efficiency of 3.4 \% at 1 MeV. 

\begin{figure}[hht]  
   \begin{center} 
     \includegraphics[width=80mm]{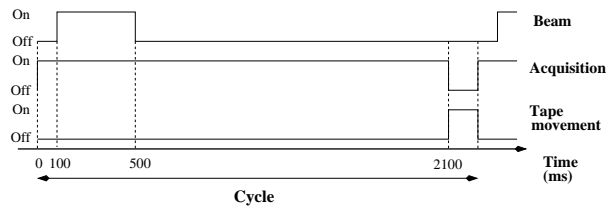} 
     \caption{Schematic view of the beam-on/beam-off cycle used in the present experiment
              together wit the acquisition cycle and the tape transport.} 
     \label{cycle}
   \end{center}
   \vspace*{-0.7cm}
\end{figure}

To provide grow-in and decay sequences of the activities, the IGISOL beam was pulsed 
(T$_{ON}$ = 400 ms, T$_{OFF}$ = 1600 ms) and the implantation tape was moved at the end of 
the beam-off period. We added a 100 ms beam off sequence just before the beam-on period 
as shown in figure~\ref{cycle}. This yielded a well-defined background and allowed for 
a clean separation of the different contributions ($^{62}$Ga, $^{62}$Cu, constant 
background) in the decay spectrum.

In order to minimize systematic effects linked to the data acquistion system, we used three 
independent acquisition systems (DAQ) which worked with different parameters (fixed dead time, 
type of records and spectra etc.), each of them running in different data record modes: 
DAQ~1 in an event-by-event mode, DAQ~2 in a cycle-by-cycle mode, and DAQ~3 in a run-by-run
mode. Energy signals for the different detectors and a time stamp were acquired 
and stored event-by-event on tape using 
the Jyv\"askyl\"a VENLA data acquisition system~\cite{Lob95} (DAQ~1). The fixed dead time 
was set to 30 $\mu$s for this data acquisition and the time binning was 1 ms/channel. 
This dead time was much longer than any other dead time from electronics modules or from
the data acquisition itself. For the other two acquisition 
systems, a coincidence between the signals of the two photomultipliers was required by a 
hardware condition. In the course of the experiment, two different fixed dead times of 
2~$\mu$s and 5~$\mu$s were tested. For the main part of the data taking, we used a fixed 
dead time of 5~$\mu$s. Again, these fixed dead times were chosen to be longer than any possible
event treatment by the electronics or data acquisition systems. The data analysis showed  
no observable difference between the different data sets. 
DAQ~2 allowed to store data cycle by cycle with 1 ms/channel. The data 
recorded for a given cycle were stored on disk while the tape was repositioned. 
DAQ~3 only acquired total grow-in/decay time spectra over a whole run (between about 200 and 
5000 cycles) with a binning of 10 ms/channel. Table \ref{DAQ} summarizes the different 
characteristics of each DAQ.

\begin{table}[hht] 
\begin{center}
{\footnotesize
\begin{tabular}{ccccc}
DAQ   & Fixed & Time    & Running & Recorded \\
      & dead time    & binning & mode    & data \\
\hline\\
N$^{\circ}$1 & 30 $\mu$s  & 1 ms/ch  & listmode & time distribution\\
 &  &  &  & $\beta$ energy spectra\\
 &  &  &  & $\gamma$ energy spectra\\
 &  &  &  & counting rate\\
N$^{\circ}$2 & 5 $\mu$s & 1 ms/ch  & cycle & time distribution\\
N$^{\circ}$3 & 5 $\mu$s & 10 ms/ch & run & time distribution\\
\end{tabular}
\caption{Characteristics of the three independent data acquisition systems used 
         during the experiment.}  
\label{DAQ}}
\end{center}
\end{table}

\section{Half-life measurements}  

The first paragraph of this section will be devoted to the way we did the analysis 
for each acquisition system. In the second part, we will investigate the possible 
systematic error sources as well as the influence of the different contaminants 
and finally we will present the half-life results.

\subsection{Analysis with three acquisition systems}

As mentioned above, DAQ~1 stored the data in an event-by-event mode. 
As in addition to the time distribution also the energy signals from the 
photomultipliers were written on tape, software conditions allowed to remove the noise 
contribution to the events stored. Different cuts were applied; however, no influence 
on the half-life could be observed. 

As the data were written in listmode, the beam-on/beam-off cycles could be reconstructed
offline and time spectra were obtained for each cycle. Therefore, the data from DAQ~1 and 
DAQ~2 could be analysed on a cycle-by-cycle basis which will be described shortly in the
following. 

Before accepting a cycle for further analysis, different checks were performed: 
i) the counting rate for the cycle had to be higher than a user defined limit (we tested
different limits in order to make sure that the counting rate limit does not alter the 
half-life. We finally used 20 counts per cycle as the limit).
ii) A fit of the time distribution of each cycle was performed with a function taking into
account the different contributions to the time spectrum. This fit had to yield a reduced
$\chi^2$ value of better than 5 (other values were tested as well, but no influence could 
be found on the half-life).

Cycles which passed these two selection criteria were then corrected channel-by-channel
for the dead-time. For this purpose, each channel content was divided by a correction factor 
of $f_{dt} = 1/(1+t_{dt}R)$ where $t_{dt}$ is the fixed dead-time and $R$ is the counting rate
per second in each channel. After this correction, the cycles were summed up to yield the
time spectrum for a given run.

In order to make sure that the data analysis does not introduce any bias to the half-life, 
a set of simulation data was generated for each accepted cycle using the same counting rate
for the different contributions ($^{62}$Ga, $^{62}$Cu, constant background), as determined
during the cycle fitting mentioned above. For the half-life, fixed values were used for
$^{62}$Ga (116.2~ms) and for $^{62}$Cu (584.4~s). We should mention that due to its very 
long half-life (9.186 h) compared to both $^{62}$Ga and $^{62}$Cu, we always considered the 
$^{62}$Zn contribution as a constant background. These simulated data were subject
to the same dead-time as the fixed dead-time used experimentally (30~$\mu$s and 5~$\mu$s
for DAQ~1 and DAQ~2, respectively). This procedure created a simulated data set equivalent 
to the experimental data, except that the input half-life was known. The aim was to obtain
the input half-life after treatment of the simulated data with the same programs as the
experimental data.

More details about the selection criteria, the fitting procedure and the generation of
the simulated data sets can be found in reference~\cite{Bla04}.

The data acquisition DAQ~3 stored all the data for a given run. Therefore, no correction 
for the fixed dead-time was possible. However, as could be shown with the data from  
DAQ~2, the dead-time correction, for a fixed dead-time as short as 5$\mu$s and the counting 
rate of the present experiment (less than 1000/s), had only
a minor influence (0.08ms) on the half-life determined. Assuming the same change in half-life
for DAQ~3, we will corrected the half-life from DAQ~3 by this time and add it quadartically
to the error for DAQ~3.

For the whole body of data, the further analysis was then very similar. In a first fit with fixed
half-lives for $^{62}$Ga and $^{62}$Cu, the fit parameters for the counting rates of the
different contributions were determined. This fit was performed over the whole time range of 
the cycles. A second fit over the whole time range was then performed with the $^{62}$Ga
half-life as an additional free parameter. The final fit was performed on the decay part
of the spectrum only, with three free parameters: the $^{62}$Ga rate and half-life as well
as the background rate. The parameters describing the $^{62}$Cu contribution
were kept at the values from the previous fit. The fact that we fixed the $^{62}$Cu contribution
does not influence the fit results in any significant way. In fact, due to its rather 
long half-life (584.4~s), the $^{62}$Cu contribution can basically be treated as a constant background
in the decay part of the time distribution (see below). This third fit yielded the final half-life value
for each run and each data acquisition system.

\subsection{Systematic errors and possible influences of contaminants}

As we performed part of the fits on the whole time range of the cycles, the beam-on and beam-off
times are crucial for a good description of the experimental spectra by the fit. In order to study
the influence of the exact time stamp when the beam was switched on or off, we performed fits
to the data with fixed switching times (100~ms and 500~ms after the start of a cycle) and with 
variable switching times, where the times were treated as fit parameters. As the quality of the
fit and thus the resulting half-life depended strongly on these times, we finally used fixed 
values, however, as adjusted in the analysis where they were treated as free parameters.

An important correction to the half-life comes from the dead-time. In order to study the influence
of this correction, we performed tests with different fixed dead-times generated by hardware
for DAQ~1 and DAQ~2. The half-life values determined for $^{62}$Ga for fixed dead-times of 2$\mu$s and 5$\mu$s
were in nice agreement. In addition, DAQ~3 had a much longer dead-time of 30$\mu$s. After correction 
we obtained half-life values in agreement with DAQ~1 and DAQ~2 for each run.

An independent check of the dead-time correction as well as of other parts of the analysis 
procedure was performed with the simulated data mentioned above. As shown below, after analysis
of these data the input half-life was nicely reproduced.

Another check performed on the data was to start the fit later in the time spectrum.
If the spectra were affected by a significant dead-time, a fit from the beginning of the decay
curve should yield a longer half-life than a fit where the first part of the decay-time distribution
is excluded. Such a behaviour could be explained by the fact that the first part of the time 
distribution is more subject to dead-time due to its higher counting rate than
a later part in the time distribution. However, no effect was observed.

As the IGISOL system is not chemically selective, all mass A=62 nuclei produced in the reaction
are transported to the detection station. This means for example that much more $^{62}$Cu and 
$^{62}$Zn was deposited on the collection tape. However, as mentioned above, due to its rather 
long half-life $^{62}$Zn is not of concern. In contrary, $^{62}$Cu has to be considered as its
production rate was about 250 times higher than the one of $^{62}$Ga. Therefore we tested different
procedures in order to take into account the $^{62}$Cu distribution. It turned out that this 
contaminant could be treated as a constant contamination in the decay part of the spectrum.
In fact, whatever we did with the $^{62}$Cu contribution, the fit adjusted the level of the 
background in the opposite way, as it is expected for a basically constant contribution from
$^{62}$Cu.

Finally, the fact that consistent results have been obtained for the three acquisition systems 
indicates that acquistion related phenomena like the dead-time were treated correctly.

\subsection{Half-life results}

\begin{figure}[ht] 
   \begin{center} 
    \includegraphics[width=80mm]{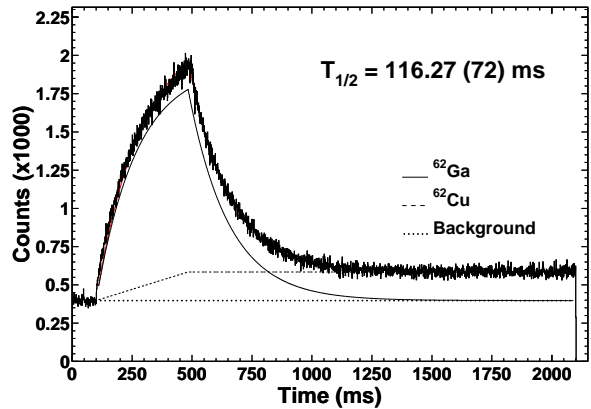} 
    \caption{Experimental data and fit for one run. The fit was performed with two exponential 
             components plus a constant background. The lines show the different contributions of
             $^{62}$Ga, $^{62}$Cu and the background. The fits yields a half-life of
             116.27(72) ms. It corresponds to about 6$\times$10$^5$ $^{62}$Ga detected for this particular run.}
    \label{run14}
   \end{center}
\end{figure}

Approximatively 10$^7$ $\beta$ decay events of $^{62}$Ga were re\-corded. Figure \ref{run14} 
presents the experimental data and the result of the fit for one run together with its 
extracted half-life.

\begin{figure}[hht] 
   \begin{center} 
    \includegraphics[width=80mm,angle=0]{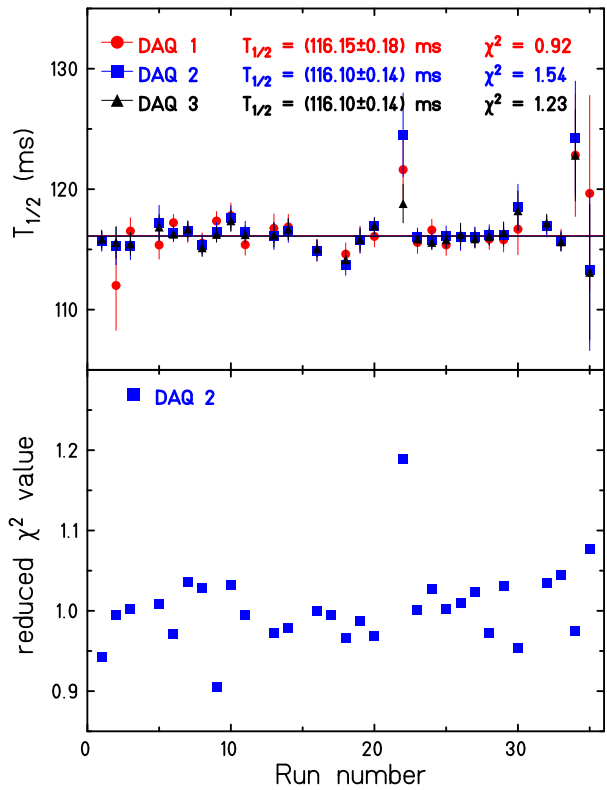} 
    \caption{Upper part: Experimental half-lives determined for the different runs and for the three data 
             acquisition systems. The figure gives also the error-weighted average for each data
             set as well as the reduced $\chi^2$ value for the averaging procedure. Here the error bars
             are not yet normalised with the square root of the reduced $\chi^2$ value.
             Lower part: Reduced $\chi^2$ values for the half-life fits of the individual runs.
             Only the reduced $\chi^2$ values for the second acquisition are shown as an example.
            }
    \label{results}
   \end{center}
\end{figure}

The half-life determined for each run and each acquisition system as well as the reduced $\chi^2$
value for each individual fit are shown in figure~\ref{results}. The half-lives scatter around
116~ms and the reduced $\chi^2$ values are close to unity except for one low-statstics run. 
The error-weigh\-ted average yields the final half-life for each data
acqui\-stion system and its associated analysis. However, due to the fact that the averaging 
procedure does not give an acceptable reduced $\chi^2$ value for two of the data 
acquisition systems, we normalized the error bars of the individual half-lives 
for each run by the square root of the reduced $\chi^2$ value, as proposed by 
the Particle Data Group~\cite{Hag02}. The new average obtained with the 
increased error bars yields the final result for each data set. These results are given in 
table~\ref{results_bis}. As our final result we adopt the error-weighted average of these
three results. However, as the data for the three acquisition systems originate from the 
same data sample, we keep as the final error the smallest error bar of the three data 
acquisitions. This result is therefore 116.09(17)~ms.
It will be compared below to other experimental half-lives.

\begin{table}[tth] 
\begin{center}
{\footnotesize
\begin{tabular}{cccc}
DAQ 1 & DAQ 2 & DAQ 3 & Final result\\
\hline\\
116.15(18) & 116.10(17) & 116.02(17) & 116.09(17)\\
\end{tabular}
\caption{Half-lives determined for each acquisition system. These values are already corrected for
         the reduced $\chi^2$ value of the averaging procedure for the different runs. 
         For DAQ~3, we corrected for the dead time as discussed above and increased the error accordingly.
         The last column corresponds to the error-weighted average value of the three
         acquisition systems together with the smallest of the individual error bars.}  
\label{results_bis}}
\end{center}
\end{table}

\begin{figure}[hht] 
   \begin{center} 
    \includegraphics[height=80mm,angle=-90]{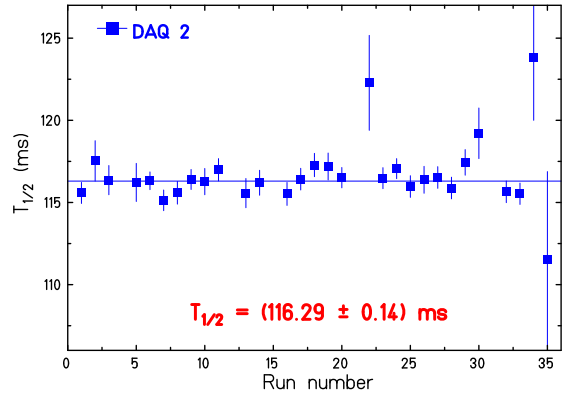} 
    \caption{Results obtained with the simulation data analysed in the same way as 
             the experimental data. As an example we display the results for acquistion 2.
             For an input half-life of 116.2~ms, a half-life value of 
             116.29(14)~ms was obtained.}
    \label{sim}
   \end{center}
\end{figure}

The simulation data which were generated with the same parameters as the experimental data
and treated in the same way as described above are shown in figure~\ref{sim}. The dead-time corrected
data yield an average half-life of 116.29(14)~ms for an input value of 116.2~ms.

\section{Branching ratios}  

Another goal of the experiment described here was to measure branching ratios for all 
important decay branches, in particular for allowed Gamow-Teller bran\-ches and non-analog 
Fermi transitions, with the Fermi transition to the first excited 0$^+$ state 
in $^{62}$Zn located at 2.33 MeV being the most important transition to measure. If observed, 
the branching ratio for the $0^+_1 \rightarrow 0^+_2$ Fermi transition allows for 
a direct comparison to theoretical prediction of one part of the isospin-mixing correction 
$\delta_c$.

\begin{figure}[hht] 
   \begin{center} 
    \includegraphics[width=80mm]{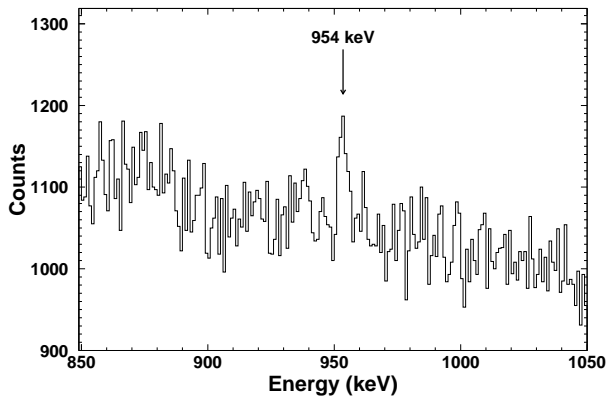} 
    \caption{Gamma-energy spectrum from the germanium detector array. The 954~keV  line 
             corresponding to the de-excitation of the first excited 2$^+$ state to the 
             ground state in $^{62}$Zn is indicated.}
    \label{g954}
   \end{center}
\end{figure}

Figure \ref{g954} shows the $\beta$-gated $\gamma$-ray spectra in the vinicity of the 954 keV $\gamma$ ray 
from the decay of the first excited 2$^+$ state in $^{62}$Zn. After efficiency correction and 
normalisation to the number of $^{62}$Ga implanted, we find a $\gamma$ branching ratio for this 
line of (0.11 $\pm$ 0.04) \%.

According to selection rules, the $2^+$ state can not be fed directly by an allowed 
$\beta$ decay. The explanation is that the $2^+$ state acts as a collector state
which collects most of the $\beta$-decay strength from higher-lying $0^+$ and $1^+$ levels.
Therefore, we searched also for $\gamma$-$\gamma$ coincidences with the 954 keV. However, no 
clear indication for a $\gamma$ cascade was found with $\gamma$-ray energies below 2.8 MeV. 
In particular, we looked for a 1376 keV $\gamma$ transition from the first excited 0$^+$ to 
the first excited 2$^+$ state. However, neither the $\gamma$-$\gamma$ coincidence data with 
the 954 keV line nor a direct analysis of the singles spectrum at 1376~keV yielded any 
evidence for such a line. From the direct search, we established an upper limit of 
0.039 \%  (95 \% confidence level) for the branching ratio for this transition by 
fitting the germanium spectra with a Gaussian plus a linear background term.

\begin{table*}[bht] 
\begin{center}
{\footnotesize
\begin{tabular}{ccccccc}
Alburger~\cite{Alb78} & Chiba et al.~\cite{Chi78} & Davids et al.~\cite{Dav79} & Hyman et al.~\cite{Hym03} & Blank et al.~\cite{Bla04} & Present work & Average \\
\hline\\
115.95(30) ms & 116.4(15) ms & 116.34(35) ms & 115.84(25) ms & 116.19(4) ms & 116.09(17) ms & 116.175(38) ms \\
\end{tabular}
\caption{Experimental results of half-life measurements for $^{62}$Ga. The error-weighted 
         average value of all measurements is presented in the last column.}
\label{moyenne}}
\end{center}
\end{table*}

\section{Discussion and conclusions}

\begin{figure}[th] 
   \begin{center} 
    \includegraphics[width=80mm]{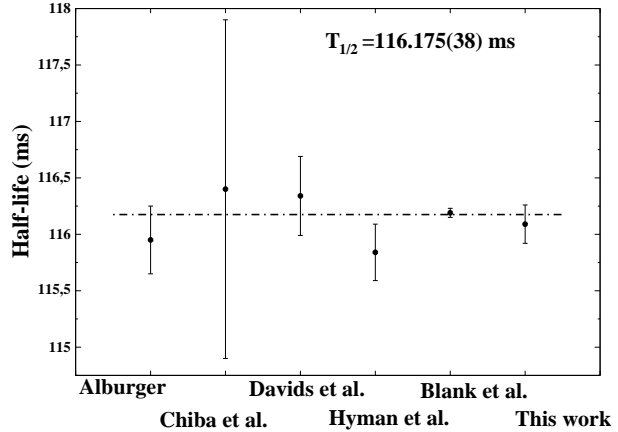} 
    \caption{Experimental half-life measurements for $^{62}$Ga. The horizontal line 
             represents the error-weighted average value yielding T$_{1/2}$ = 116.175(38)~ms.}
    \label{moyenne_bis}
   \end{center}
\end{figure}

Our experimental result for the half-life of $^{62}$Ga of T$_{1/2}$ = 116.09(17) ms is in nice agreement 
with all previous half-life measurements (see table~\ref{moyenne} and figure~\ref{moyenne_bis}). 
It is the second most precise value ever measured. The fact that all experimental half-lives 
are in agreement gives some confidence that the half-life of $^{62}$Ga is known now with 
sufficient precision in order to serve for the determination of the corrected $Ft$ value 
for the super-allowed $\beta$ decay of $^{62}$Ga. By averaging the experimentally measured 
half-life values, we obtain the new recommended half-life of $^{62}$Ga of 116.175(38)~ms.

Several $\beta$-decay branching ratio measurements of $^{62}$Ga have been  
published recently~\cite{Bla02,Hym03,Dor02} and they are all in excellent agreement with our results.
As mentioned above, we find that the intensity of the 954 keV line, de-exciting the first excited 2$^+$
state to the ground state of $^{62}$Zn, is 0.11(4) \%. In table~\ref{tab:BR}, we compare our result for
this decay branch with the other experimental data published.

Based on his results for the branching ratios in conjunction with shell-model calculations,
Hyman {\it et al.}~\cite{Hym03} determined a branching ratio of 99.8(1) \% for the super-allowed $\beta$ decay
from the ground state of $^{62}$Ga to the ground state of $^{62}$Zn, a result which is not altered
by our new results.

\begin{table*}[!bht] 
\begin{center}
{\footnotesize
\begin{tabular}{ccccccc}
Blank~\cite{Bla02} & Hyman et al.~\cite{Hym03} & D\"oring et al.~\cite{Dor02} & Present work & Average \\
\hline\\
0.12(3) \% & 0.120(21) \% & 0.106(17) \% & 0.11(4) \% & 0.112(11) \% \\
\end{tabular}
\caption{Experimental results of branching-ratio measurements for the decay of the first excited
         $2^+$ in $^{62}$Ga to the ground state. The error-weighted 
         average value of all measurements of 0.112(11)\% is presented in the last column.}
\label{tab:BR}}
\end{center}
\end{table*}

Similar agreement is obtained for the upper limit for the feeding of the excited 0$^+$ state at 2.33 MeV
fed by a forbidden Fermi transition. Our limit of 0.039 \% (95\% confidence limit) compares well with
a limit of 0.017\%~\cite{Bla02} and 0.043\%~\cite{Hym03} (one-sigma limits) obtained earlier. 
In a two-level mixing model, this upper limit can be compared to the part of the isospin correction 
$\delta_c$ which is due to level mixing~\cite{Har98}. From our new limit of 0.039\%, we calculate
an upper limit of the mixing matrix element of $\delta_{IM}$ = 0.21 which can 
be compared to theoretical results from Ormand and Brown of $\delta_{IM}$ = 0.169 and 
0.079~\cite{Orm95} using two different $fp$ shell interactions 
and from Towner and Hardy of $\delta_{IM}$ = 0.085(20)\cite{Tow02}. Our upper limit is still in
agreement with these predictions, which means that more precise experimental data for the 
feeding of the $0^+_2$ state in $^{62}$Zn are needed to distinguish between the different 
model predictions.

\begin{figure}[hht] 
   \begin{center} 
    \includegraphics[width=80mm]{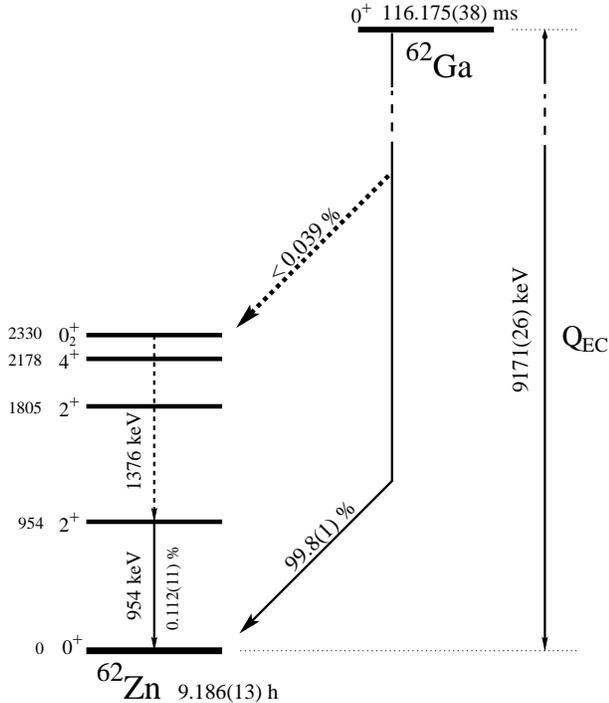} 
    \caption{$\beta$ decay scheme of $^{62}$Ga indicating the Q value, the half-life
             as well as the branching ratios.}
    \label{decay}
   \end{center}
\end{figure}

Figure \ref{decay} presents a partial decay scheme for $^{62}$Ga which summarizes the different branching 
ratios, the half-life, and the $\beta$-decay Q value. These values will be used to determine the $ft$ value for
the super-allowed $\beta$ decay of $^{62}$Ga.

The experimental results now available for the $\beta$ decay of $^{62}$Ga, 
i.e. the Q value of 9171(26) keV~\cite{Dav79}, the half-life of 116.175(38) ms, and
the branching ratio of 99.8(1)\% yield a {\it ft} value of {\it ft}=3056(47) s. Finally, 
by taking into account the correction factors as calculated by Towner and Hardy~\cite{Tow02}, we 
derive a corrected {\it Ft} value of {\it Ft}=3057(47) s. This value is basically the same as 
the one for {\it ft}, as the correction factors nearly compensate each other.

This result agrees well with the average {\it Ft} value, 3072.2(8) s, obtained from the nine 
well-known superallowed decays of nuclei with 10$\leq$A$\leq$54. However, the poor precision 
on Q$_{EC}$ and to some extent on the branching ratio severely limits the precision one can 
achieve for the moment on the {\it Ft} value for $^{62}$Ga. Mass measurement campains on 
$^{62}$Ga and $^{62}$Zn are planned or have recently started at ISOLTRAP at CERN and at the new 
JYFL-trap in Jyv\"askyl\"a to reach a precision for both systems of the order of 10$^{-8}$. 
More experiments on the precise measurement of the branching ratios are planned for the near 
future. Therefore, a high-precision {\it Ft} value can be expected in the near future for $^{62}$Ga.
These results together with results recently published for $^{74}$Rb~\cite{Kel04} will allow us to test the CVC 
hypothesis through $\beta$ decay of higher-Z nuclei, to verify the isospin correction terms more 
carefully, and to yield a better understanding of the weak-interaction standard model.

\section*{Acknowledgment}
The authors would like to acknowledge the continous effort of the whole 
Jyv\"askyl\"a accelerator laboratory staff for ensuring a smooth running 
of the experiment. This work was supported in part by the Conseil R\'egional d'Aquitaine
and by the European Union Fifth Framework Programme "Improving Human Potential -
Access to Research Infrastructure" (Contract No. HPRI-CT-1999-00044).
We also acknowledge support from the Academy of Finland under the Finnish
Centre of Excellence Programme 2000-2005 (Project No. 44875, Nuclear and
Condensed Matter Physics Programme at JYFL).

\end{document}